\documentclass[12pt,twocolumn]{article}
\usepackage{epsfig}

\begin{document}

\begin{titlepage}
\begin{center}
{\Large Generalized information and entropy measures in physics}

\vspace{2.cm} {\bf Christian Beck}

\vspace{0.5cm}

School of Mathematical Sciences, Queen Mary, University of London,

Mile End Road, London E1 4NS, UK

\vspace{2cm}

\end{center}

\abstract{The formalism of statistical mechanics can be
generalized by starting from more general measures of information
than the Shannon entropy and maximizing those subject to suitable
constraints. We discuss some of the most important examples of
information measures that are useful for the description of
complex systems. Examples treated are the R\'{e}nyi entropy, Tsallis
entropy, Abe entropy, Kaniadakis entropy, Sharma-Mittal
entropies, and a few more. Important concepts such as the
axiomatic foundations, composability and Lesche stability of
information measures are briefly discussed.
Potential applications in physics include complex
systems with long-range interactions and metastable states,
scattering processes in particle physics, hydrodynamic
turbulence, defect turbulence, optical lattices, and quite generally
driven nonequilibrium systems with fluctuations of temperature.}

\vspace{1.3cm}

\end{titlepage}

\newpage

\small

\section{How to measure information}

\subsection{Prologue}

How should one measure information?
There is no unique answer to
this. There are many different information measures, and what
measure of information is the most suitable one will in general
depend on the problem under consideration. Also, there are
different types of information. For example, the information a
reader gets from reading a book on quantum field theory is
different from the one he gets from reading Shakespeare's {\em Romeo
and Juliet}. In general one has to distinguish between elementary
and advanced information concepts. The elementary information is
just related to technical details such as, for example, the
probability to observe certain letters in a long sequence of
words. The advanced information is related to the information the
reader really gets out of reading and understanding a given text,
i.e. this concept requires coupling to a very complex system such
as the brain of a human being.

\begin{figure}
\epsfig{file=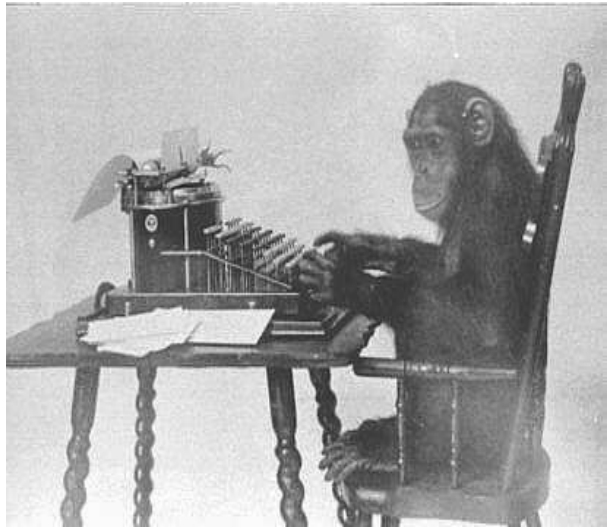, width=8cm, height=7cm} \caption{There is no
obvious way to measure the information contents of given symbol
sequences. While it is relatively easy to distinguish between a
random sequence of symbols and Shakespeare's {\em Romeo and Juliet} in
terms of suitable elementary information measures, it is less
obvious how to distinguish the fact that the advanced information
contents given by Shakespeare's {\em Romeo and Juliet} is different
from the one given by a book on quantum field theory.}
\end{figure}

In physics, the missing information on the concrete state of a
system is related to the {\em entropy} of the system. Entropy is
an elementary information concept. Many different physical
definitions of entropy can be given, and what makes up a
`physically relevant entropy' is often subject to `heated'
discussions. Misunderstandings with respect to the name `entropy'
seem to be the rule rather than the exception within the past 130
years. Generally one may use the name `entropy' as a synonym for a
possible quantity to measure missing information, keeping in mind
that large classes of possible functions will potentially do the
job, depending on application.

The entire formalism of statistical
mechanics can be regarded as being based on maximizing the entropy
($=$ missing information)
of the system under consideration subject to suitable constraints,
and hence naturally the question arises how to
measure this missing information in the first place \cite{BS}.
While normally
one chooses the Shannon information measure,
in principle more general information
measures (that contain the Shannon information as a special case)
can be chosen as well. These then
formally lead to generalized versions of statistical
mechanics when they are maximized \cite{tsallis, mendes, kaniadakis, naudts, abc, chavanis2}.

In this paper we describe some
generalized information and entropy measures that are useful in this context.
We discuss their most important properties, and point out potential
physical applications.
The physical examples we
choose are the statistics of cosmic rays \cite{cosmic},
defect turbulence \cite{daniels}, and optical lattices
\cite{lutz, renzoni}, but the general techniques developed have applications for a variety
of other complex systems as well,
such as
driven nonequilibrium systems with
large-scale fluctuations of
temperature (so-called
superstatistical systems \cite{beck-cohen,BCS}), hydrodynamic turbulence
\cite{reynolds, prl}, scattering processes in particle physics
\cite{curado, e+e-}, gravitationally interacting systems \cite{plastino, chavanis}
and Hamiltonian systems with long-range interactions and metastable states
\cite{rapisarda,raptsa}. There are applications outside physics as well,
for example
in mathematical finance \cite{borland},
biology \cite{bio} and medicine \cite{chen}.

\subsection{Basic concepts}

One usually restricts the concept of an information measure to an
information that is a function of a given probability distribution
of events (and nothing else)\footnote{An exception to this rule is
the Fisher information, which depends on gradients of the
probability density but will not be discussed here.}. The basic
idea is as follows. Consider a sample set of $W$ possible events.
In physics events are often identified as possible microstates of
the system. Let the probability that event $i$ occurs be denoted
as $p_i$. One has from normalization
\begin{equation}
\sum_{i=1}^Wp_i=1.
\end{equation}
We do not know which event will occur. But suppose that one of
these events, say $j$, finally takes place. Then we have clearly
gained some information, because before the event occurred we did
not know which event would occur.

Suppose that the probability $p_j$ of that observed event $j$ is
close to 1. This means we gain very little information by the
observed occurrence of event $j$, because this event was very
likely anyway. On the other hand, if $p_j$ is close to zero, then
we gain a lot of information by the actual occurrence of event
$j$, because we did not really expect this event to happen. The
information gain due to the occurrence of a single event $j$ can
be measured by a function $h(p_j)$, which should be close to zero
for $p_j$ close to 1. For example, we could choose $h(p_j)=\log
p_j$, the logarithm to some suitable basis $a$. If this choice of
$a$ is $a=2$ then $h$ is sometimes called a 'bit-number'
\cite{BS}. But various other functions $h(p_j)$ are possible as
well, depending on the application one has in mind. In other
words, an information measure should better be regarded as a
man-made construction useful for physicists who don't fully
understand a complex system but try to do so with their limited
tools and ability. Once again we emphasize that an information
measure is not a universally fixed quantity. This fact has led to
many misunderstandings in the community.

In a long sequence of independent trials, in order to determine an
average information gain by the sequence of observed events $i$ we
have to weight the information gain associated with a single event
with the probability $p_i$ that event $i$ actually occurs. That is
to say, for a given function $h$ the average information gained
during a long sequence of trials is
\begin{equation}
I(\{p_i\})=\sum_{i=1}^Wp_ih(p_i). \label{BBB}
\end{equation}
Many information measures studied in the literature are indeed of
this simple trace form. But other forms are possible as well.
One then defines the entropy $S$ as `missing
information', i.e.\
\begin{equation}
S=-I.
\end{equation}
This means the entropy is defined as our missing information on
the actual occurrence of events, given that we only know the
probability distribution of the events.

If the probability distribution is sharply peaked around one
almost certain event $j$, we gain very little information from our
long-term experiment of independent trials: The event $j$ will
occur almost all of the time, which we already knew before.
However, if all events are uniformly distributed, i.e.\ $p_i=1/W$
for all $i$, we get a large amount of information by doing this
experiment, because before we did the experiment we had no idea
which events would actually occur, since they were all equally
likely. In this sense, it is reasonable to assume that an
(elementary) information measure should take on an extremum
(maximum or minimum, depending on sign) for the uniform
distribution. Moreover, events $i$ that cannot occur ($p_i=0$) do
not influence our gain of information in the experiment at all.
In this way we arrive at the most basic principles an information
measure should satisfy.

\subsection{The Khinchin axioms}

There is a more formal way to select suitable (elementary)
information measures, by formulating a set of axioms and then
searching for information measures that satisfy these axioms. A
priori, there is an infinite set of possible information
measures, not only of the simple form (\ref{BBB}) but of more
general forms as well, based on arbitrary functions of the entire
set of $p_i$. How can we select the most suitable ones, given
certain requirements we have in mind? Of course, what is `most
suitable' in this context will in general depend on the
application we have in mind. The most appropriate way of dealing
with this problem is to postulate some basic and essential
properties the information measure one is interested in should
have, and then to derive the functional form(s) that follows from
these postulates.

Khinchin \cite{khinchin} has formulated four axioms that describe
the properties a `classical' information measure $I$ should
have (by `classical' we mean an information measure yielding
ordinary Boltzmann-Gibbs type of statistical mechanics):

\subsubsection*{Axiom 1}
\begin{equation}
I=I(p_1,\cdots ,p_W)
\end{equation}
That is to say, the information measure $I$ only depends on the
probabilities $p_i$
of the events and nothing else.

\subsubsection*{Axiom 2}
\begin{equation}
I(W^{-1},\ldots ,W^{-1})\leq  I(p_1,\cdots ,p_W)
\end{equation}
This means the information measure $I$ takes on an absolute
minimum for the uniform distribution $(W^{-1},\ldots ,W^{-1})$,
any other probability distribution has an information contents
that is larger or equal to that of the uniform distribution.

\subsubsection*{Axiom 3}
\begin{equation}
I(p_1,\ldots,p_W)=I(p_1,\ldots,p_W,0)
\end{equation}
This means the information measure $I$ should not change if the
sample set of events is enlarged by another event that has
probability zero.

\subsubsection*{Axiom 4}
\begin{equation}
I(\{p_{ij}^{I, II}\})=I(\{p_i^I\})+\sum_i p_i^I  I(\{p^{II}(j|i)
\})
\end{equation}
This axiom is slightly more complicated and requires a longer
explanation. The axiom deals with the composition of two systems I
and II (not necessarily independent). The probabilities of the
first system are $p_i^I$, those of the second system are
$p_j^{II}$. The joint system I,II is described by the joint
probabilities $p_{ij}^{I, II}=p_i^Ip^{II}(j|i)$, where
$p^{II}(j|i)$ is the conditional probability of event $j$ in
system II under the condition that event $i$ has already occurred in
system $I$. $I(\{p^{II}(j|i)\})$ is the conditional information of
system II formed with the conditional probabilities $p^{II}(j|i)$,
i.e.\ under the condition that system I is in state $i$.

The meaning of the above axiom is that it postulates that the
information measure should be independent of the way the
information is collected. We can first collect the information in
the subsystem II, assuming a given event $i$ in system I, and then
sum the result over all possible events $i$ in system I, weighting
with the probabilities $p_i^I$.

For the special case that system I and II are independent the
probability of the joint system factorizes as
\begin{equation}
p_{ij}^{I,II}=p_i^Ip_j^{II},
\end{equation}
and only in this case, axiom 4 reduces to the rule of additivity
of information for independent subsystems:
\begin{equation}
I(\{ p_{ij}^{I,II}\})=I(\{p_i^I\})+I(\{p_j^{II}\}) \label{indep}
\end{equation}

Whereas there is no doubt about Axioms 1--3, the reader
immediately notices that Axiom 4 requires a much longer
explanation. From a physical point of view, Axiom 4 is a much less
obvious property. Why should information be independent from the
way we collect it?

To illustrate this point, we may consider a simple example of an
information-collecting system, a first-year undergraduate student
trying to understand physics. This student will learn much more
if he first attends a course on classical mechanics, collecting
all available information there, and then attends a course on
quantum mechanics. If he does it the other way round, he will
probably hardly understand anything in the course on quantum
mechanics, since he does not have the necessary prerequisites. So
attending the quantum mechanics course first leads to zero
information gain.
Apparently, the order in which the information of the two courses
(the two subsystems) is collected is very important and leads to
different results in the achieved knowledge.

In general complex systems, the order in which information is
collected can be very relevant. This is a kind of information
hysteresis phenomenon. In these cases we have situations where
the replacement of Axiom 4 by something more general makes
physical sense. We will come back to this in section 3.

\subsection{The Shannon entropy}

It is easy to verify that the celebrated Shannon entropy, defined
by
\begin{equation}
S=-k \sum_{i=1}^Wp_i\ln p_i
\end{equation}
satisfies all four of the Khinchin axioms. Indeed, up to an
arbitrary multiplicative constant, one can easily show (see, e.g.,
\cite{BS}) that this is the only entropic form that satisfies all
four Khinchin axions, and that it follows uniquely (up to a
multiplicative constant) from these postulates. $k$ denotes the
Boltzmann constant, which in the remaining sections will be set
equal to 1. For the uniform distribution, $p_i=1/W$, the Shannon
entropy takes on its maximum value
\begin{equation}
S=k\ln W,
\end{equation}
which is Boltzmann's famous formula, carved on his grave in
Vienna.
\begin{figure}
\epsfig{file=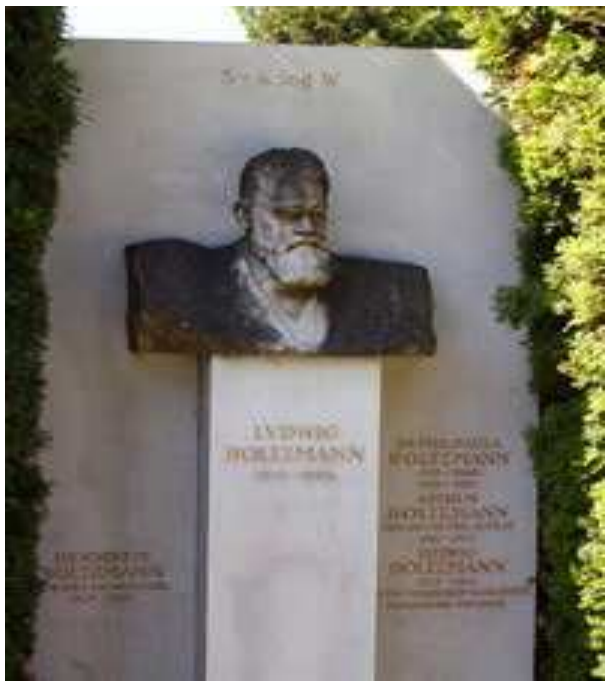, width=8cm, height=9cm} \caption{The grave
of Boltzmann in Vienna. On top of the gravestone the formula $S=k
\log W$ is engraved. Boltzmann laid the foundations for
statistical mechanics, but his ideas were not widely accepted
during his time. He comitted suicide in 1906.}
\end{figure}
 Maximizing the Shannon entropy subject to suitable
constraints leads to ordinary statistical mechanics (see section
4.2). In thermodynamic equilibrium, the Shannon entropy can be
identified as the `physical' entropy of the system, with the
usual thermodynamic relations. Generally, the Shannon entropy has
an enormous range of applications not only in equilibrium
statistical mechanics but also in coding theory, computer
science, etc.

It is easy to verify that $S$ is a concave function of the
probabilities $p_i$, which is an important property to formulate
statistical mechanics. Remember that concavity of a differentiable
function $f(x)$ means $f''(x)\leq 0$ for all $x$. For the Shannon entropy
one has
\begin{eqnarray}
\frac{\partial}{\partial p_i}S&=&-\ln p_i -1 \\
\frac{\partial^2}{\partial p_i\partial
p_j}S&=&-\frac{1}{p_i}\delta_{ij}\leq 0,
\end{eqnarray}
and hence, as a sum of concave functions of the single probabilities $p_i$,
the Shannon entropy $S$ is a concave function.

In classical mechanics, one often has a continuous variable $u$
with some probability density $p(u)$, rather than discrete
microstates $i$ with probabilities $p_i$. In this case the
normalization condition reads $\int_{-\infty}^\infty p(u)du=1$,
and the Shannon entropy associated with this probability density
is defined as
\begin{equation}
S=-\int_{-\infty}^{\infty}du p(u)\ln (\sigma p(u)), \label{14}
\end{equation}
where $\sigma$ is a scale parameter that has the same dimension as
the variable $u$. For example, if $u$ is a velocity (measured in
units of $m/s$), then $p(u)$, as a probability density of velocities,
has the dimension $s/m$, since $p(u)du$ is a dimensionless quantity.
As a consequence, one needs to introduce the scale parameter $\sigma$
in eq.~(\ref{14}) to make the argument of the logarithm dimensionless.

Besides the Shannon information, there are lots of other
information measures. We will discuss some of the most important
examples in the next section. Some information measures are more
suitable than others for the description of various types of
complex systems. We will discuss the axiomatic foundations that
lead to certain classes of information measures. Important
properties to check for a given information measure are convexity,
additivity, composability, and stability. These properties can
help to select the most suitable generalized information measure
to describe a given class of complex systems.

\section{More general information measures}

\subsection{The R\'{e}nyi entropies}

We may replace axiom 4 by the less stringent condition
(\ref{indep}), which just states that the entropy of independent
systems should be additive. In this case one ends up with other
information measures which are called the {\em R\'{e}nyi entropies}.
\cite{renyi}. These are defined for an arbitrary real parameter
$q$ as
\begin{equation}
S_q^{(R)} =\frac{1}{q-1}\ln \sum_i p_i^q.\label{renyi}
\end{equation}
The summation is over all events $i$ with $p_i\not=0$. The R\'{e}nyi
entropies satisfy the Khinchin axioms 1--3 and the additivity
condition (\ref{indep}). Indeed, they follow uniquely from these
conditions, up to a multiplicative constant. For $q\to 1$ they
reduce to the Shannon entropy:
\begin{equation}
\lim_{q\to 1}S_q^{(R)}=S,
\end{equation}
as can be easily derived by
setting $q=1+\epsilon$ and doing a perturbative expansion in the small
parameter $\epsilon$ in eq.~(\ref{renyi}).

The R\'{e}nyi information measures are important for the
characterization of multifractal sets (i.e., fractals with a
probability measure on their support \cite{BS}), as well as for
certain types of applications in computer science. But do they
provide a good information measure to develop a generalized
statistical mechanics for complex systems?

At first sight it looks nice that the R\'{e}nyi entropies are additive
for independent subsystems for general $q$, just as the Shannon
entropy is for $q=1$. But for non-independent subsystems I and II
this simplicity vanishes immediately: There is no simple formula
of expressing the total R\'{e}nyi entropy of a joint system as a
simple function of the R\'{e}nyi entropies of the interacting
subsystems.

Does it still make sense to generalize statistical mechanics using
the R\'{e}nyi entropies? Another problem arises if one checks whether
the R\'{e}nyi entropies are a convex function of the probabilities.
The R\'{e}nyi entropies do not possess a definite convexity---the
second derivative with respect to the $p_i$ can be positive or
negative. For formulating a generalized statistical mechanics,
this poses a serious problem. Other generalized information
measures are better candidates--we will describe some of those
in the following.



\subsection{The Tsallis entropies}
The {\em Tsallis entropies} (also called $q$-entropies) are given by the
following expression \cite{tsallis}:
\begin{equation}\label{s}
S_q^{(T)} =  \frac{1}{q-1}\left(1-\sum_{i=1}^W p_i^q\right).
\label{tsallis-entro}
\end{equation}
One finds definitions similar to eq.~(\ref{tsallis-entro}) already
in earlier papers such as e.g.\ \cite{havrda}, but it was Tsallis
in his seminal paper \cite{tsallis} who for the first time
suggested to generalize statistical mechanics using these
entropic forms. Again $q\in\cal{R}$ is a real parameter, the
entropic index. As the reader immediately sees, the Tsallis
entropies are different from the R\'{e}nyi entropies: There is no
logarithm anymore. A relation between R\'{e}nyi and Tsallis entropies
is easily derived by writing
\begin{equation}
\sum_ip_i^q=1-(q-1)S_q^{(T)}=e^{(q-1)S_q^{(R)}}
\end{equation}
which implies
\begin{equation}
S_q^{(T)}=\frac{1}{q-1}(1-e^{(q-1) S_q^{(R)}}). \label{tsallis-e}
\end{equation}
Apparently the Tsallis entropy is a monotonous function of the
R\'{e}nyi entropy, so any maximum of the Tsallis entropy will also be
a maximum of the R\'{e}nyi entropy and vice versa. But still, Tsallis
entropies have many distinguished properties that make them a
better candidate for generalizing statistical mechanics than,
say, the R\'{e}nyi entropies.

One such property is concavity. One easily verifies that
\begin{eqnarray}
\frac{\partial}{\partial p_i}S_q^{(T)}&=&-\frac{q}{q-1}p_i^{q-1} \\
\frac{\partial^2}{\partial p_i\partial p_j}S_q^{(T)}&=&
-qp_i^{q-2}\delta_{ij}.
\end{eqnarray}
This means that, as a sum of concave functions, $S_q^{(T)}$ is concave for all $q>0$ (convex for all
$q<0$). This property is missing for the R\'{e}nyi entropies. Another
such property is the so-called Lesche-stability, which is
satisfied for the Tsallis entropies but not satisfied by the
R\'{e}nyi entropies (see section 3.3 for more details).

The Tsallis entropies also contain the Shannon entropy
\begin{equation}
S= -\sum_{i=1}^Wp_i\ln p_i\label{s1}
\end{equation}
as a special case. Letting $q\to 1$ we have
 \begin{equation}
S_1^{(T)}=\lim_{q\to 1}S_q^{(T)} =S
\end{equation}

As expected from a good information measure, the Tsallis entropies
take on their extremum for the uniform distribution $p_i =
1/W\;\;\forall{i}$. This extremum is given by
\begin{equation}
S_q^{(T)}= \frac{W^{1-q}-1}{1-q}
\end{equation}
which, in the limit $q \rightarrow 1$, reproduces Boltzmann's
celebrated formula $S= \ln W$.

It is also useful to write down the definition of Tsallis
entropies for a continuous probability density $p(u)$ with
$\int_{-\infty}^\infty p(u)du=1$, rather than a discrete set of
probabilities $p_i$ with $\sum_i p_i=1$. In this case one defines
\begin{equation}
S_q^{(T)}=\frac{1}{q-1}\left(
1-\int_{-\infty}^{+\infty}\frac{du}{\sigma} (\sigma p(u))^q\right)
, \label{25}
\end{equation}
where again $\sigma$ is a scale parameter that has the same dimension as
the variable $u$. It is introduced
for a similar reason as before, namely to make the
integral in eq.~(\ref{25}) dimensionless
so that it can be substracted from 1.
For $q\to 1$ eq.~(\ref{25}) reduces to the Shannon entropy
\begin{equation}
S_1^{(T)}=S=-\int_{-\infty}^{\infty}du p(u) \ln (\sigma p(u)).
\end{equation}


A fundamental property of the Tsallis entropies is the fact that
they are not additive for independent subsystems. In fact, they
have no chance to do so, since they are different from the R\'{e}nyi
entropies, the only solution to eq.~(\ref{indep}).

To investigate this in more detail, let us consider two
independent subsystems I and II with probabilities $p_i^I$ and
$p_j^{II}$, respectively. The probabilities of joint events $i,j$
for the combined system I,II are $p_{ij}=p_i^Ip_j^{II}$. We may
then consider the Tsallis entropy for the first system, denoted as
$S_q^I$, that of the second system, denoted as $S_q^{II}$, and
that of the joint system, denoted as $S_q^{I,II}$. One has
\begin{equation}
S_q^{I, II}=S_q^I +S_q^{II}-(q-1) S_q^{I}S_q^{II}.\label{pa}
\end{equation}

 {\small
{\bf Proof of eq.~(\ref{pa}):} We may write
\begin{eqnarray}
\sum_i p_i^{Iq} &=&1-(q-1)S_q^I \label{aa}\\ \sum_j p_j^{IIq}
&=&1-(q-1)S_q^{II} \label{bb}
\\
\sum_{i,j} p_{ij}^q&=&
\sum_i(p_i^{I})^q\sum_j(p_j^{II})^q  \nonumber
\\ \,&=&
1-(q-1)S_q^{I, II}.\label{there}
\end{eqnarray}
From eqs.~(\ref{aa}) and (\ref{bb}) it also follows that
\begin{eqnarray}
\sum_i(p_i^{I})^q\sum_j(p_j^{II})^q &=&1-(q-1)S_q^I-(q-1)S_q^{II}
\nonumber \\
&+&(q-1)^2 S_q^IS_q^{II}. \label{here}
\end{eqnarray}
Combining eqs.~(\ref{there}) and (\ref{here}) one ends up with
eq.~(\ref{pa}), q.e.d.}

Apparently, if we put together two independent subsystems then the
Tsallis entropy is not additive but there is a correction term
proportional to $q-1$, which vanishes for $q=1$ only, i.e.\ for
the case where the Tsallis entropy reduces to the Shannon entropy.
Eq.~(\ref{pa}) is sometimes called the `pseudo-additivity'
property.

\begin{figure}
\epsfig{file=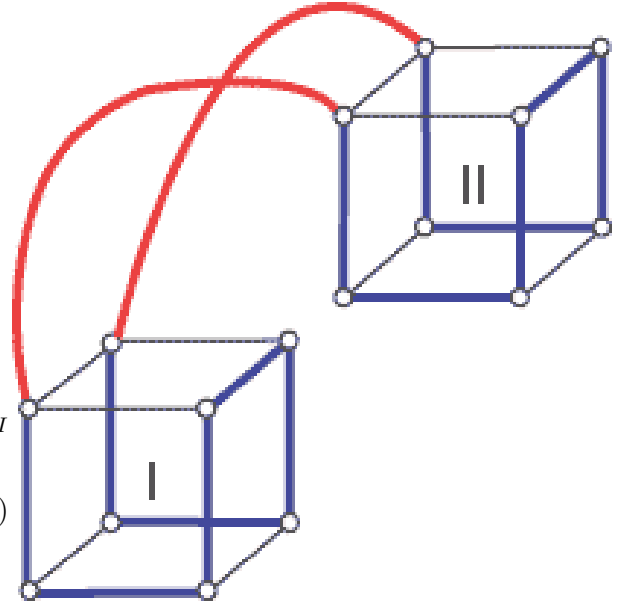, width=8cm, height=8cm}
\caption{If the nonadditive entropies $S_q$ are used to measure
information, then
the information contents of
two systems I, II (blue) that are put together is not equal
to the sum of the
information contents of the isolated single systems.
In other words, there is
always an interaction between the subsystems (red).}
\end{figure}

Eq.~(\ref{pa}) has given rise to the name {\em nonextensive
statistical mechanics}. If we formulate
a generalized statistical mechanics based on maximizing Tsallis
entropies, then the (Tsallis) entropy of {\em independent} systems
is not additive. However, it turns out that for special types of
{\em correlated} subsystems, the Tsallis entropies do become
additive if the subsystems are put together \cite{gell-mann}.
This means, for these types of correlated complex systems a
description in terms of Tsallis entropies  in fact can make
things simpler as compared to using the Shannon entropy, which is
non-additive for correlated subsystems.

\subsection{Landsberg-Vedral entropy}

Let us continue with a few other examples of generalized
information measures. Consider
\begin{equation}
S_q^{(L)}=\frac{1}{q-1}\left(\frac{1}{\sum_{i=1}^Wp_i^q}-1\right).
\end{equation}
This measure was studied by Landsberg and Vedral
\cite{landsberg-vedral}. One immediately sees that the
Landsberg-Vedral entropy is related to the Tsallis entropy
$S_q^{(T)}$ by
\begin{equation}
S_q^{(L)}=\frac{S_q^{(T)}}{\sum_{i=1}^Wp_i^q},
\end{equation}
and hence $S_q^{(L)}$ is sometimes also called {\em normalized Tsallis
entropy}. $S_q^{(L)}$ also contains the Shannon entropy as a
special case
\begin{equation}
\lim_{q\to 1}S_q^{(L)}=S_1
\end{equation}
and one readily verifies that it also satisfies a
pseudo-additivity condition for independent systems, namely
\begin{equation}
S_q^{(L)I,II}=S_q^{(L)I}+S_q^{(L)II}+(q-1)S_q^{(L)I}S_q^{(L)II}
.
\end{equation}
This means that in the pseudo-additivity relation (\ref{pa}) the role of
$(q-1)$ and $-(q-1)$ is exchanged.

\subsection{Abe entropy}

Abe \cite{abe-entropy} introduced a kind of symmetric modification
of the Tsallis entropy, which is invariant under the exchange $q
\longleftrightarrow q^{-1}$. This is given by
\begin{equation}
S_q^{Abe}=-\sum_i \frac{p_i^q-p_i^{q^{-1}}}{q-q^{-1}} \label{sabe}
\end{equation}
This symmetric choice in $q$ and $q^{-1}$ is inspired by the
theory of quantum groups which often exhibits invariance under
the `duality transformation' $q\to q^{-1}$. Like Tsallis entropy,
the Abe entropy is also concave. In fact, it is related to the
Tsallis entropy $S_q^T$ by
\begin{equation}
S_q^{Abe}= \frac{(q-1)S_q^T-(q^{-1}-1)S_{q^{-1}}^T}{q-q^{-1}}.
\end{equation}
Clearly the relevant  range of $q$ is now just the unit
interval$(0,1]$, due to the symmetry $q \longleftrightarrow
q^{-1}$: Replacing $q$ by $q^{-1}$ in eq.~(\ref{sabe}) does not
change anything.

\subsection{Kaniadakis entropy}
The {\em Kaniadakis entropy} (also called $\kappa$-entropy) is defined
by the following expression \cite{kaniadakis}
\begin{equation}
S_\kappa = \sum_i \frac{p_i^{1+\kappa}-p_i^{1-\kappa}}{2\kappa}
\end{equation}
Again this is a kind of deformed Shannon entropy, which reduces
to the original Shannon entropy for $\kappa =0$. We also note that
for small $\kappa$, and by  writing $q=1+\kappa$, $q^{-1} \approx
1-\kappa$, the Kaniadakis entropy approaches the Abe entropy.
Kaniadakis was motivated to introduce this entropic form by
special relativity: The relativistic sum of two velocities of
particles of mass $m$ in special relativity satisfies a similar
relation as the Kaniadakis entropy does, identifying
$\kappa=1/mc$.
Kaniadakis entropies are also concave and Lesche stable (see
section 3.3).

\subsection{Sharma-Mittal entropies}

These are two-parameter families of entropic forms
\cite{sharma-mittal}. They can be written in the form
\begin{equation}
S_{\kappa ,r}= -\sum_i p_i^r\left( \frac{p_i^\kappa -p_i^{-\kappa}}{2 \kappa} \right)
\end{equation}
Interestingly, they contain many of the entropies mentioned so far as special cases.
The Tsallis entropy is obtained for $r=\kappa$ and $q=1-2\kappa$. The Kaniadakis entropy
is obtained for $r=0$. The Abe entropy is obtained for $\kappa =\frac{1}{2} (q-q^{-1})$
and $r=\frac{1}{2}(q+q^{-1})-1$. The Sharma-Mittal entropes are concave and Lesche stable.

\section{Selecting a suitable information measure}

\subsection{Axiomatic foundations}

The Khinchin axioms apparently are the right axioms to obtain the
Shannon entropy in a unique way, but this concept may be too
narrow-minded if one wants to describe general complex systems.
In physics, for example, one may be interested in nonequilibrium
systems with a stationary state, glassy systems, long transient
behaviour in systems with long-range interactions, systems with
multifractal phase space structure etc. In all these cases one
should be open-minded to allow for generalizations of axiom 4,
since it is this axiom that is least obvious in the given
circumstances.

Abe \cite{abe-axiom} has shown that the Tsallis entropy follows
uniquely (up to an arbitrary multiplicative constant) from the
following generalized version of the Khinchin axioms. Axioms 1--3
are kept, and Axiom 4 is replaced by the following more general
version:

\subsubsection*{New Axiom 4}
\begin{equation}
S_q^{I, II}=S_q^I+S_q^{II|I}-(q-1)S_q^IS_q^{II|I}
\end{equation}
Here $S_q^{II|I}$ is the conditional entropy formed with the
conditional probabilities $p(j|i)$ and averaged over all states
$i$ using the so-called escort distributions $P_i$:
\begin{equation}
S_q^{II|I}=\sum_i P_i S_q(\{ p(j|i) \}).
\end{equation}
Escort distributions $P_i$ were introduced quite generally in \cite{BS}
and are defined for any given probability distribution $p_i$ by
\begin{equation}
P_i= \frac{p_i^q}{\sum_i p_i^q}.
\end{equation}
For $q=1$, the new axiom 4 reduces to the old Khinchin axiom 4,
i.e. $S_q^{I, II}=S_q^I+S_q^{II|I}$. For independent systems I and
II, the new axiom 4 reduces to the pseudo-additivity property
(\ref{pa}).

The meaning of the new axiom 4 is quite clear. It is a kind of
minimal extension of the old axiom 4: If we collect information
from two subsystems, the total information should be the sum of
the information collected from system I and the conditional
information from system II, plus a correction term. This
correction term can a priori be anything, but we want to restrict
ourselves to information measures where
\begin{equation}
S^{I,II}=S^I+S^{II|I}+g(S^I,S^{II|I}),
\end{equation}
where $g(x,y)$ is some function. The property that the entropy of
the composed system can be expressed as a function of the
entropies of the single systems is sometimes referred to as the
{\em composability} property. Clearly, the function $g$ must depend
on the entropies of {\em both} subsystems, for symmetry reasons.
The simplest form one can imagine is that it is given by
\begin{equation}
g(x,y)=\mbox{const} \cdot xy,
\end{equation}
i.e.\ it is proportional to both the entropy of the first system
and that of the second system. Calling the proportionality
constant $q-1$, we end up with the new axiom 4.

It should, however, be noted that we may well formulate other
axioms, which then lead to other types of information measures.
The above generalization is perhaps the one that requires least
modifications as compared to the Shannon entropy case. But
clearly, depending on the class of complex systems considered, and
depending on what properties we want to describe, other axioms
may turn out to be more useful. For example, Wada and Suyari
\cite{wada} have suggested a set of axioms that uniquely lead to
the Sharma-Mittal entropy.

\subsection{Composability} Suppose we have a given complex system
which consists of subsystems that interact in a complicated way.
Let us first analyze two subsystems I and II in an isolated way
and then put these two dependent systems I and II together. Can
we then express the generalized information we have on the total
system as a simple function of the information we have on the
single systems? This question is sometimes referred to as the
composability problem.

The Tsallis entropies are composable in a very simple way. Suppose
the two systems I and II are not independent. In this case
one can still write the joint probability $p_{ij}$ as a product
of the single probability $p_i$ and conditional probability
$p(j|i)$, i.e.\ the probability of event $j$ under the condition
that event $i$ has already occurred is
\begin{equation}
p_{ij}=p(i|j)p_j.
\end{equation}
The conditional Tsallis entropy associated with system II (under
the condition that system I is in state $i$) is given by
\begin{equation}
S_q^{II|i}=\frac{1}{q-1}(1-\sum_j p(j|i)^q).
\end{equation}
One readily verifies the relation
\begin{equation}
S_q^I+\sum_ip_i^qS_q^{II|i}=S_q^{I, II}. \label{rela}
\end{equation}
This equation is very similar to that satisfied by the Shannon
entropy in Axiom 4. In fact, the only difference is that there is
now an exponent $q$ that wasn't there before. It means our
collection of information is biased: Instead of weighting the
events in system I with $p_i$ we weight them with $p_i^q$. For
$q=1$ the above equation of course reduces to the fourth of the
Khinchin axioms, but only in this case. Hence, for general
$q\not=1$, the Tsallis information is {\em not} independent of the
way it is collected for the various subsystems.

To appreciate the simple composability property of the Tsallis
entropy, let us compare with other entropy-like functions, for
example the R\'{e}nyi entropy. For the R\'{e}nyi entropy there is no
simple composability property similar to eq.~(\ref{rela}). Only
the exponential of the Renyi entropy satisfies a relatively simple
equation, namely
\begin{equation}
\exp \left((q-1)S_q^{(R)I,II}\right)=\sum_ip_i^q\exp \left( (q-1)
S_q^{(R)II|i}\right) . \label{corres}
\end{equation}
However, by taking the exponential one clearly removes the
logarithm in the definition of the R\'{e}nyi entropies in
eq.~(\ref{renyi}). This means one is effectively back to the
Tsallis entropies.

\subsection{Lesche stability}

Physical systems contain noise. A necessary requirement for a
generalized entropic form $S[p]$ to make physical sense is that
it must be stable under small perturbations. This means a small
perturbation of the set of probabilities $p:=\{p_i\}$ to a new
set $p'=\{p_i'\}$  should have only a small effect on the value
$S_{max}$ of $S_q[p]$ in the thermodynamic state that maximizes
the entropy. This should in particular be true in the limit $W\to
\infty$ (recall that $W$ denotes the number of microstates). The
stability condition can be mathematically expressed as follows
\cite{lesche}:

\subsubsection*{Stability condition}
For every $\epsilon >0$ there is a $\delta >0$ such that
\begin{equation}
||p-p'||_1\leq \delta \Longrightarrow \left|
\frac{S[p]-S[p']}{S_{max}}\right| < \epsilon \label{stability}
\end{equation}
for arbitrarily large $W$. Here $||A||_1=\sum_{i=1}^W|A_i|$
denotes the $L_1$ norm of an observable $A$.

Abe \cite{abe-pre} has proved that the Tsallis entropies are
Lesche-stable, i.e.\ they satisfy eq.~(\ref{stability}) for all
$q$, whereas the R\'{e}nyi entropies and the Landsberg entropies are
not stable for any $q\not= 1$ (for a discrete
set of probabilities $p_i$ with $W\to \infty$).
This is an important criterion to
single out generalized entropies that may have potential physical
meaning. According to the stability criterion, the Tsallis
entropies are stable and thus may be associated with physical
states, whereas the other two examples of entropic forms have a
stability problem.
Kaniadakis entropies and Sharma Mittal
entropies are also Lesche-stable. Only entropies that are
Lesche-stable are good candidates for physically relevant
information measures. For a recent re-investigation of
these types of stability problems in a physical setting, see \cite{thurner}.

\section{Maximizing generalized entropies}

\subsection{A rigorous derivation of statistical mechanics?}

The rigorous foundations of statistical mechanics are a kind of
miracle. There is little progress in rigorously deriving
statistical mechanics from the microscopic classical Hamiltonian
equations of motion, neither there is a rigorous derivation
starting from quantum mechanics or quantum field theory.
It is
almost surprising how well statistical mechanics
works in daily life, given its lack of rigorous
derivation from other microscopic theories.

The problem is that for a rigorous derivation of statistical mechanics
from dynamical systems theory one needs
the underlying dynamical
system to be {\em ergodic}, and even that is not enough: It
should have the
stronger property of {\em mixing}.
Ergodicity essentially means
that
typical trajectories fill out the entire phase space
(which implies that for typical trajectories the time average
is equal to the ensemble average) and mixing
means asymptotic independence, i.e.\ the correlation function
of distant events decays to zero if the time difference
between the events goes to infinity. For strongly
chaotic dynamical systems
(i.e.\
those exhibiting exponential sensitive dependence
on the initial conditions) one normally expects the mixing
property to hold (though there are some mathematical subtleties
here).
From a mathematical point of view,
the mixing property is the theoretical ingredient
that is needed to guarantee the approach to
an equilibrium state in statistical mechanics.

Unfortunately,
ergodicity and mixing can only be rigorously
proved for simple toy examples
of dynamical systems, for example the discrete-time map $x_{n+1}=1-2x_n^2$
with initial values in the interval $[-1,1]$ or other
very simple toy models (see, e.g. \cite{BS}). For realistic systems
of physical relevance, such as the Hamiltonian equations of a large
number of weakly or strongly interacting particles, a rigorous
mathematical proof of the mixing property does
not exist, and the deeper reason why
statistical mechanics works
so well in typical situations remains a miracle.

\subsection{Jaynes' information theory}
In view of the fact that there are no rigorous foundations of
statistical mechanics, one usually sticks to some simple principle such
as the maximum entropy principle in order to `derive' it.
Jaynes has given a simple and plausible interpretation of the
maximum entropy principle \cite{jaynes}. His interpretation is
purely based on concepts from information theory, and applicable
to many problems, not only to equilibrium statistical mechanics.

In simple words, the idea is as follows. Assume we have only
limited information on a system containing many particles or
constituents. We may know the mean values of some observables
$M^\sigma$, $\sigma=1,\cdots, s$ but nothing else. For example,
we may just know one such quantity, the mean energy of all
particles and nothing else ($s=1$). What probability
distributions $p_i$ should we now assume, given that we have such
limited information on the system?

Suppose we measure information with some information measure
$I(\{p_i\})=:I[p]$. Among all distributions possible that lead to
the above known mean values $M^\sigma$ we should select those that
do not contain any unjustified prejudices. In other words, our
information measure for the relevant probabilities should take on
a minimum, or the entropy ($=$ negative information) should take a
maximum, given the constraints. For, if the information associated
with the selected probability distribution does not take on a
minimum, we have more information than the minimum information,
but this means we are pre-occupied by a certain belief or
additional information, which we should have entered as a
condition of constraint in the first place.

%

Of course, if we have no knowledge on the system at all ($s=0$),
the principle yields the uniform distribution $p_i=1/W,
\;i=1,\ldots W$ of events. For this to happen, the information
measure $I[p]$ must only satisfy the second Khinchin axiom,
nothing else. In statistical mechanics, the corresponding ensemble
is the microcanonical ensemble.

If some constraints are given, we have to minimize the information
($=$ maximize the entropy) subject to the given constraints. A
constraint means that we know that some observable $\tilde{M}$ of
the system, which takes on the values $M_i$ in the microstates
$i$, takes on the fixed mean value $M$. In total, there can be $s$
such constraints, corresponding to $s$ different observables
$\tilde{M}^\sigma$:
\begin{equation}
\sum_i p_i M_i^\sigma =M^\sigma  \;\;\;\;(\sigma=1, \ldots , s).
\end{equation}
For example, for the canonical ensemble of equilibrium statistical
mechanics one has the constraint that the mean value $U$ of the
energies $E_i$ in the various microstates is fixed:
\begin{equation}
\sum_i p_i E_i =U
\end{equation}
We may also regard the fact that the probabilities $p_i$ are
always normalized as a constraint obtained for the special choice
$\tilde{M}=1$:
\begin{equation}
\sum_i p_i =1.
\end{equation}

To find the distributions that maximize the entropy under the
given constraints
one can use the method of Lagrange
multipliers. One simply defines a function $\Psi [p]$ which is the
information measure under consideration plus the condition of
constraints multiplied by some constants $\beta_\sigma$ (the
Lagrange multipliers):
\begin{equation}
\Psi[p]=I[p]+\sum_\sigma \beta_\sigma (\sum_i p_iM_i^\sigma).
\label{psi}
\end{equation}
One then looks for the minimum of this function in the space of
all possible probabilities $p_i$. In practice, these distributions
$p_i$ are easily obtained by evaluating the condition
\begin{equation}
\frac{\partial}{\partial p_i} \Psi[p]=0 \;\;\;(i=1,\ldots ,W),
\end{equation}
which means that $\Psi$ has a local extremum. We obtain
\begin{equation}
\frac{\partial}{\partial p_i} I[p] +\sum_\sigma \beta_\sigma
M_i^\sigma = 0 \label{nursery}
\end{equation}
All this is true for {\em any} information measure $I[p]$, it need not be
the Shannon information.
At this point we see why it is important that the information
measure $I[p]$ is convex: We need a well-defined inverse function
of $\frac{\partial}{\partial p_i}I[p]$, in order to uniquely solve
eq.~(\ref{nursery}) for the $p_i$. This means
$\frac{\partial}{\partial p_i}I[p]$ should be a monotonous
function, which means that $I[p]$ must be convex.

Note that Jaynes' principle is (in principle) applicable to all
kinds of complex systems, many different types of observables, and
various types of information measures. There is no reason to
restrict it to equilibrium statistical mechanics only. It's
generally applicable to all kinds of problems where one has
missing information on the actual microscopic state of the system
and wants to make a good (unbiased) guess of what is happening
and what should be done. The concept of avoiding unjustified
prejudices applies in quite a general way. An important question
is which information measure is relevant for which system.
Clearly, the Shannon entropy is the right information measure to
analyse standard type of systems in equilibrium statistical
mechanics. But other systems of more complex nature can
potentially be described more effectively if one uses different
information measures, for examples those introduced in the
previous section.

\subsection{Ordinary statistical mechanics}

For ordinary statistical mechanics, one has $I[p]=\sum_i p_i\ln
p_i$ and $\frac{\partial}{\partial p_i} I[p]=1+\ln p_i$. For the
example of a canonical ensemble eq.~(\ref{psi}) reads
\begin{equation}
\Psi[p] =\sum_i p_i\ln p_i +\alpha \sum_i p_i + \beta \sum_i p_i
E_i
\end{equation}
and eq.~(\ref{nursery}) leads to
\begin{equation}
\ln p_i +1+\alpha +\beta E_i =0.
\end{equation}
Hence the maximum entropy principle leads to the canonical
distributions
\begin{equation}
p_i=\frac{1}{Z}e^{-\beta E_i}.
\end{equation}
The partition function $Z$ is related to the Lagrange multiplier
$\alpha$ by
\begin{equation}
Z:=\sum_i e^{-\beta E_i} =e^{1+\alpha}.
\end{equation}

\subsection{Generalized statistical mechanics}

More generally we may start from a generalized information
measure of the trace form
\begin{equation}
I[p]=-S[p]=\sum_i p_i h(p_i)
\end{equation}
where $h$ is some suitable function, as introduced before. Tsallis
entropy, Abe entropy, Kaniadakis entropy, Sharma-Mittal entropy
and Shannon entropy are examples that can all be cast into this
general form, with different functions $h$ of course. Again let
us consider the canonical ensemble (the extension to further
constraints/other ensembles is straightforward). The functional
to be maximized is then
\begin{equation}
\Psi [p]= \sum_i p_i h(p_i) +\alpha \sum_i p_i +\beta \sum_i p_i
E_i
\end{equation}
By evaluating the condition
\begin{equation}
\frac{\partial}{\partial p_i} \Psi [p]=0
\end{equation}
we obtain
\begin{equation}
h(p_i)+p_ih'(p_i)+\alpha +\beta E_i =0
\end{equation}
Defining a function $g$ by
\begin{equation}
g(p_i):=h(p_i)+p_ih'(p_i)
\end{equation}
we end up with
\begin{equation}
g(p_i)=-\alpha -\beta E_i
\end{equation}
Hence, if a unique inverse function $g^{-1}$ exists, we have
\begin{equation}
p_i=g^{-1} (-\alpha -\beta E_i)
\end{equation}
and this is the generalized canonical distribution of the
generalized statistical mechanics.


Let us consider a few examples of interesting functions functions $h$.
For the Shannon entropy one has of course
\begin{equation}
h(p_i)=\ln p_i.
\end{equation}
For the Tsallis entropy,
\begin{equation}
h(p_i)= \frac{p_i^{q-1}-1}{q-1}=: \log_{2-q} (p_i).
\end{equation}
This is like a deformed logarithm that approaches the ordinary
logarithm for $q\to 1$. In fact, a useful definition commonly
used in the field is the so-called $q$-logarithm defined by
\begin{equation}
\log_q (x):= \frac{x^{1-q}-1}{1-q}.
\end{equation}
Its inverse function is the $q$-exponential
\begin{equation}
e_q^x:= (1+(1-q)x)^{\frac{1}{1-q}}.
\end{equation}
For the Kaniadakis entropy one has
\begin{equation}
h(p_i)= \frac{p_i^\kappa -p_i^{-\kappa}}{2\kappa} =: \ln_\kappa (x),
\end{equation}
where the $\kappa$-logarithm is defined as
\begin{equation}
\ln_\kappa (x) =\frac{x^\kappa -x^{-\kappa}}{2 \kappa}.
\end{equation}
Its inverse is the $\kappa$-exponential
\begin{equation}
exp_\kappa (x) = ( \sqrt{1+\kappa^2 x^2}+\kappa x)^{\frac{1}{\kappa}}
\end{equation}
Essentially, the generalized canonical distributions obtained by
maximizing Tsallis entropies are given by $q$- exponentials
of the energy $E_i$ and
those by maximizing Kaniadakis entropies are
$\kappa$-exponentials. Both decay with a power law for large
values of the energy $E_i$.

\subsection{Nonextensive statistical me\-chanics}

Let us consider in somewhat more detail a generalized statistical
mechanics based on Tsallis entropies. If we start from the
Tsallis entropies $S_q^{(T)}$ and maximize those subject to
suitable constraint, the corresponding formalism is called
{\em nonextensive statistical mechanics}. We have
\begin{equation}
I_q^{(T)}[p]=-S_q^{(T)}[p]=\frac{1}{q-1} (1-\sum_i p_i^q),
\end{equation}
thus
\begin{equation}
\frac{\partial}{\partial p_i} I_q^{(T)}[p]=
\frac{q}{q-1}p_i^{q-1}.
\end{equation}
For a canonical ensemble eq.~(\ref{nursery}) leads to
\begin{equation}
\frac{q}{q-1} p_i^{q-1}+\alpha +\beta E_i =0.
\end{equation}
Thus the maximum entropy principle leads to generalized canonical
distributions of the form
\begin{equation}
p_i=\frac{1}{Z_q}(1-\beta (q-1)E_i)^{\frac{1}{q-1}},
\end{equation}
where $Z_q$ is a normalization constant. This is the original
formula Tsallis introduced in his paper \cite{tsallis}. These
days, however, the convention has become to replace the parameter
$q$ by
 $q'=2-q$
and then rename $q' \to q$. That is to say, the generalized
canonical distributions in nonextensive statistical mechanics are
given by the following $q$-exponentials:
\begin{equation}
p_i=\frac{1}{Z_q}(1+\beta (q-1)E_i)^{\frac{-1}{q-1}}.
\end{equation}
They live on a bounded support for $q < 1$ and exhibit power-law
decays for $q>1$.

Starting from such a $q$-generalized approach, one can easily
derive formal $q$-generalized thermodynamic relations. The details
depend a bit how the constraints on energy taken into account
\cite{mendes}. All relevant thermodynamic quantities now get an
index $q$. Typical examples of such formulas are
\begin{equation}
1/T= \beta=\partial S_q^{(T)} /\partial U_q,\;\;\forall q
\end{equation}
with
\begin{equation}
\sum_{i=1}^W (p_i)^q=(\bar{Z}_q)^{1-q},
\end{equation}
\begin{equation}
F_q \equiv U_q - T S_q = -\frac{1}{\beta}
\frac{(Z_q)^{1-q}-1}{1-q}
\end{equation}
and
\begin{equation}
U_q = -\frac{\partial}{\partial \beta}\frac{(Z_q)^{1-q}-1}{1-q},
\end{equation}
where
\begin{equation}
\frac{(Z_q)^{1-q}-1}{1-q}=\frac{(\bar{Z}_q)^{1-q}-1}{1-q}-\beta
U_q.
\end{equation}
and
\begin{equation}
\sum_{i=1}^WP_i E_i= \frac{\sum_{i=1}^W p_i^q E_i }{ \sum_{i=1}^W
p_i^q }= U_q
\end{equation}
There are some ambiguities on how to take into account the constraints,
using for example the original $p_i$ or the escort distributions $P_i$,
but we will not comment on these technicalities here.

\section{Some physical examples}

\subsection{Making contact with experimental data}

It should be clear that a direct physical measurements of
generalized entropy measures is impossible since these are
basically man-made information-theoretic tools. However, what can
be measured is the stationary probability distribution of certain
observables of a given complex system, as well as possibly some
correlations between subsystems. As we have illustrated before,
measured probability densities in some complex system that
deviate from the usual Boltzmann factor $e^{-\beta E}$ can then
be formally interpreted as being due to the maximization of a
more general information measure that is suitable as an effective
description for the system under consideration.

In this approach one regards the complex system as a kind of
`black box'. Indeed many phenomena in physics, biology, economics,
social sciences, etc.
 are so complicated that there is not a simple
equation describing them, or at least we do not know this
equation. A priori we do not know what is the most suitable way to
measure information for any output that we get from our black
box. But if a distribution $p_i$ of some observable output is
experimentally measured, we can indirectly construct a
generalized entropic form that takes a maximum for this
particular observed distribution. This allows us to make contact
with experimental measurements, make some predictions e.g. on
correlations of subsystems and translate the rather abstract
information theoretical concepts into physical reality.

\subsection{Statistics of cosmic rays}
Our first example of making contact to concrete measurements is
cosmic ray statistics \cite{cosmic}. The earth is constantly
bombarded with highly energetic particles, cosmic rays.
Experimental data of the measured cosmic ray energy spectrum are
shown in Fig.~4.
\begin{figure}
\epsfig{file=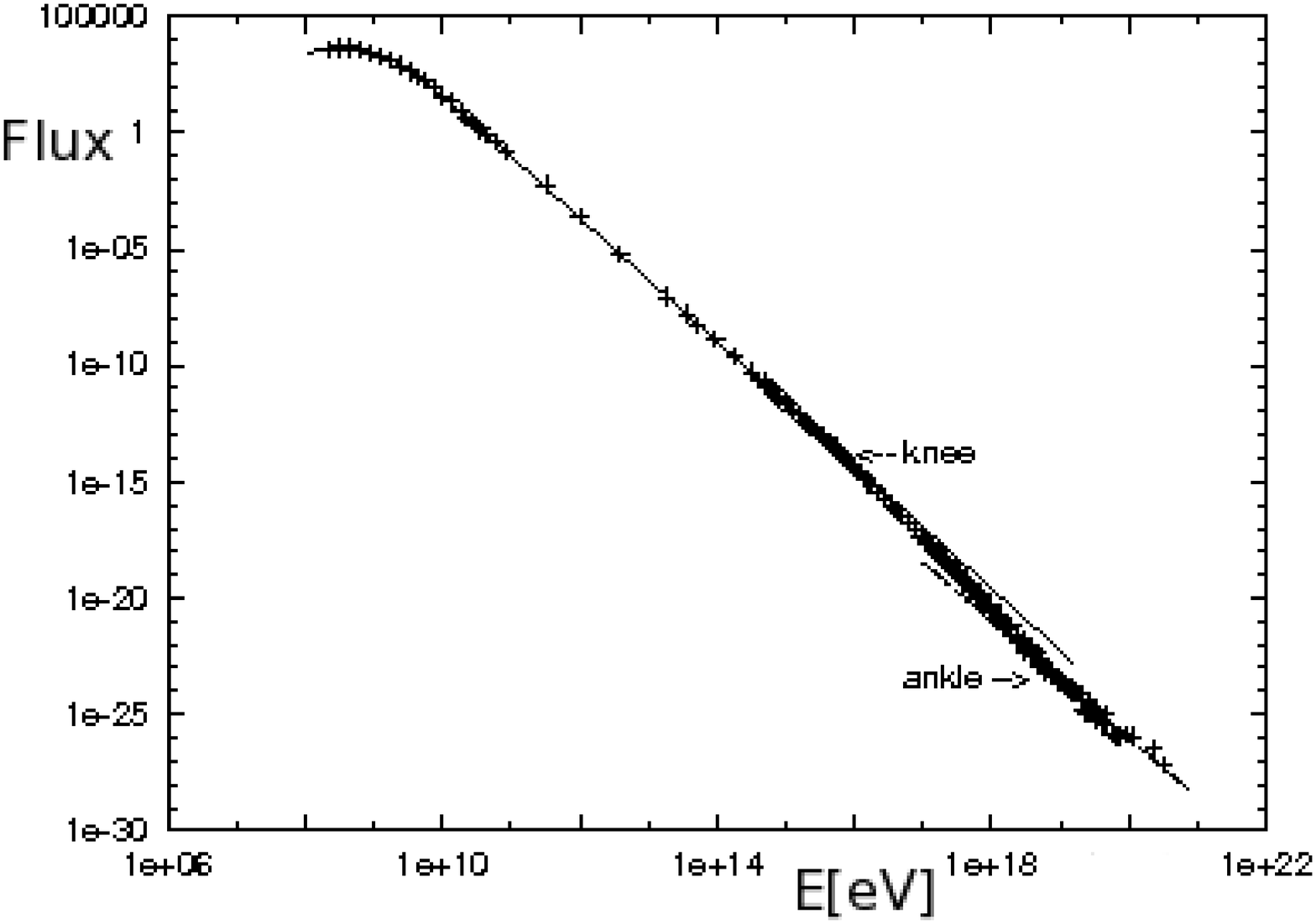, width=8cm, height=7cm}
\caption{Measured energy spectrum of cosmic rays and a fit by
eq.~(\ref{can}) with $q=1.215$.
The `knee' and `ankle' are structures that go
beyond the simple model considered here.}
\end{figure}
It has been known for a long time that the observed distribution of cosmic rays
with a given energy $E$ exhibits strongly pronounced power laws
rather than exponential decay. It turns out that the observed
distribution is very well fitted over a very large range of
energies by the formula
\begin{equation}
p(E)=C \cdot \frac{E^2}{(1+b(q-1)E)^{1/(q-1)}}. \label{can}
\end{equation}
Here $E$ is the energy of the cosmic ray particles,
\begin{equation}
E=\sqrt{c^2p_x^2+c^2p_y^2+c^2p_z^2+m^2c^4},
\end{equation}
$b=(k\tilde{T})^{-1}$ is an effective inverse temperature
variable, and $C$ is a constant representing the total flux rate.
For highly relativistic particles the rest mass $m$ can be neglected and
one has $E\approx c |\vec{p}|$. The reader immediately recognizes
the distribution (\ref{can}) as a $q$-generalized relativistic
Maxwell-Boltzmann distribution, which maximizes the Tsallis
entropy. The factor $E^2$ takes into account the available phase
space volume. As seen in Fig.~4, the cosmic ray spectrum is very
well fitted by the distribution (\ref{can}) if the entropic index
$q$ is chosen as $q=1.215$ and if the effective temperature
parameter is given by $k\tilde{T}=b^{-1}=107$ MeV.
Hence the measured cosmic ray spectrum effectively
maximizes the Tsallis entropy.

The deeper reason why this is so could be
temperature fluctuations during the production process
of the primary cosmic ray particles \cite{cosmic}. Consider quite generally
a superposition of ordinary Maxwell-Boltzmann
distributions with different inverse temperatures $\beta$:
\begin{equation}
p(E) \sim \int f(\beta) E^2 e^{-\beta E} d\beta \label{superpo}
\end{equation}
Here $f(\beta)$ is the probability density to observe
a given inverse temperature $\beta$.
If $f(\beta)$ is a Gamma distribution,
then the integration in eq.~(\ref{superpo})
can be performed and one ends up with eq.~(\ref{can})
(see \cite{cosmic} for
more details). This is the basic idea underlying
so-called superstatistical
models \cite{beck-cohen}: One does a kind
of generalized statistical mechanics where the inverse temperature $\beta$
is a random variable as well.

The effective temperature parameter $\tilde{T}$
(a kind of average temperature in the above superstatistical model)
is of the same order of magnitude
as the so-called Hagedorn temperature $T_H$ \cite{hage}, an
effective temperature well known from collider experiments. The
fact that we get from the fits something of the order of the
Hagedorn temperature is encouraging. The Hagedorn temperature is
much smaller than the center-of-mass energy $E_{CMS}$ of a typical
collision process and represents a kind of `boiling temperature'
of nuclear matter at the confinement phase transition. It is a
kind of maximum temperature that can be reached in a collision
experiment. Even largest $E_{CMS}$ cannot produce a larger
average temperature than $T_H$ due to the fact that the number of
possible particle states grows exponentially.

Similar predictions derived from nonextensive statistical
mechanics also fit measured differential cross sections in
$e^+e^-$ annihilation processes and other scattering data very
well (see e.g. \cite{curado, e+e-} for more details). The hadronic cascade
process underlying these scattering data is not well understood,
though it can be simulated by Monte Carlo simulations. If we
don't have any better theory, then the simplest model to
reproduce the measured cross sections is indeed a generalized
Hagedorn theory where the Shannon entropy is replaced by Tsallis
entropy \cite{e+e-}.

\subsection{Defect turbulence}

Our next example is so-called `defect turbulence'. Defect
turbulence shares with ordinary turbulence only the name as
otherwise it is very different. It is a phenomenon related to
convection and has nothing to do with fully developed
hydrodynamic turbulence. Consider a Rayleigh-B\'{e}nard convection
experiment: A liquid is heated from below and cooled from above.
For large enough temperature differences, interesting convection
patterns start to evolve. An inclined layer convection experiment
\cite{daniels} is a kind of Rayleigh-B\'{e}nard experiment where the
apparatus is tilted by an angle (say 30 degrees), moreover the
liquid is confined between two very narrow plates. For large
temperature differences, the convection rolls evolve chaotically.
Of particular interest are the defects in this pattern, i.e.\
points where two convection rolls merge into one (see Fig.~5).
\begin{figure}
\epsfig{file=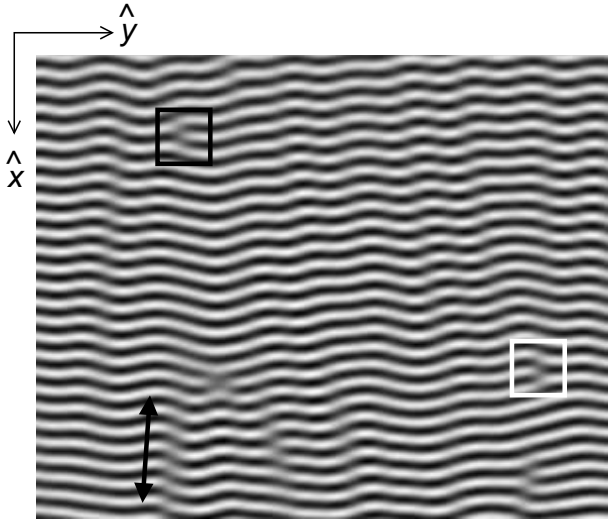, width=8cm, height=7cm}
\caption{Convection rolls and defects (black and white boxes) as
observed in the experiment of Daniels et al. \cite{daniels}}.
\end{figure}
These defects behave very much like particles. They have a
well-defined position and velocity, they are created and
annihilated in pairs, and one can even formally attribute a
`charge' to them: There are positive and negative defects, as
indicated by the black and white boxes in Fig.~5. But the theory
underlying these highly nonlinear excitations is pretty unclear,
they are like a `black box' complex system whose measured output is
velocity.

The probability density of defect velocities has been
experimentally measured with high statistics \cite{daniels}. As
shown in Fig.~6, the measured distribution is well fitted by a
$q$-Gaussian with $q \approx 1.45$.
\begin{figure}
\epsfig{file=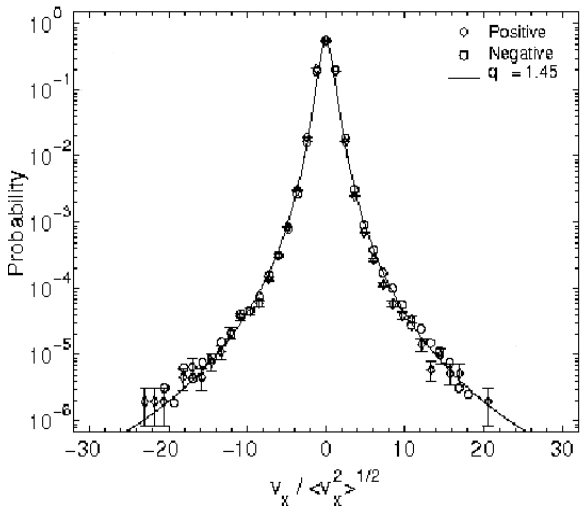, width=8cm, height=7cm}
\caption{Measured probability density of defect velocities and
fit with a $q$-Gaussian with $q=1.45$.}
\end{figure}
The defects are also observed to exhibit anomalous diffusion.
Their position $X(t)$ roughly obeys an anomalous diffusion law of
the type
\begin{equation}
\langle X^2(t) \rangle \sim t^\alpha,
\end{equation}
with $\alpha \approx 1.3$. The relation $\alpha \approx 2/(3-q)$
can be theoretically derived \cite{daniels}.

Apparently defects are a very complicated nonlinear system with
complicated interactions in a nonequilibrium environment. Their
dynamics is not fully understood so far. But we see that
effectively they seem to behave like a gas of nonextensive
statistical mechanics that leads to $q$-exponential Boltzmann
factors rather than ordinary Boltzmann factors.

\subsection{Optical lattices}

Optical lattices are standing-wave potentials obtained by
superpositions of counter-propagating laser beams. One obtains
easily tunable periodic potentials in which atoms can perform
normal and anomalous quantum transport processes. If the
potential is very deep, there is diffusive motion. If it is very
shallow, there is ballistic motion. In between, there is a regime
with anomalous diffusion that is of interest here.

Optical lattices can be theoretically described by a nonlinear
Fokker-Planck equation for the Wigner function $W(p,t)$ (the
Wigner function is an important statistical tool for the quantum
mechanical description in the phase space). The above
Fokker-Planck equation admits Tsallis statistics as a stationary
solution. This was pointed out by Lutz \cite{lutz}. The equation
is given by
\begin{equation}
\frac{\partial W}{\partial t}= - \frac{\partial}{\partial p}
[K(p)W] +\frac{\partial}{\partial p} \left[ D(p) \frac{\partial
W}{\partial p} \right]
\end{equation}
where
\begin{equation}
K(p)=-\frac{\alpha p}{1+(p/p_c)^2}
\end{equation}
is a momentum-dependent drift force and
\begin{equation}
D(p)=D_0+\frac{D_1}{1+(p/p_c)^2}
\end{equation}
a momentum-dependent diffusion constant. The stationary solution
is
\begin{equation}
W(p)= C \frac{1}{(1+\beta (q-1) E)^{\frac{1}{q-1}}}
\end{equation}
where
\begin{eqnarray}
E= &=& \frac{1}{2} p^2 \\
\beta &=& \frac{\alpha}{D_0+D_1} \\
q &=& 1+\frac{2D_0}{\alpha p_c^2}
\end{eqnarray}
So the optical lattice effectively maximizes Tsallis entropy in
its nonequilibrim stationary state.
Another way to express the entropic index in terms
of physical parameters is the formula
\begin{equation}
q=1+ \frac{44E_R}{U_0}
\end{equation}
where $E_R$ is the so-called recoil energy and $U_0$ the potential
depth. These types of $q$-exponential predictions have been
experimentally confirmed \cite{renzoni}. Lutz' microscopic theory
thus yields a theory of the relevant entropic index $q$ in terms
of system parameters.

\subsection{Epilogue}

There are many other examples of physical systems where
generalized entropies yield a useful tool to effectively describe
the complex system under consideration. Important examples
include Hamiltonian systems with long-range interactions
that exhibit metastable states
\cite{rapisarda, raptsa} as well as driven nonequilibrium systems with
large-scale fluctuations of temperature or energy dissipation, i.e.
superstatistical systems \cite{beck-cohen, wilk, prl01}. The best
way to define generalized entropies for superstatistical systems
is still subject of current research \cite{abc,
straeten,souza}. Superstatistical turbulence models yield excellent
agreement with experimental data \cite{BCS, reynolds, prl}.
Generalized statistical mechanics methods have also applications
outside physics, for example in mathematical finance
\cite{borland,bouchard}, for traffic delay statistics \cite{briggs} or in
the  medical \cite{chen} and biological sciences \cite{chavanis2}. It is often in these types of
complex systems that one does not have a concrete equation of
motion and hence is forced to do certain `unbiased guesses' on
the behaviour of the system---which for sufficiently complex
systems may lead to other entropic forms than the usual Shannon
entropy that are effectively maximized. The beauty of the
formalism is that it can be applied to a large variety of complex
systems from different subject areas, without knowing the details
of the dynamics.

\end{document}